# Facebook and the epistemic logic of friendship

Jeremy Seligman
Department of Philosophy
The University of Auckland
Auckland, New Zealand

Fenrong Liu
Department of Philosophy
Tsinghua University
Beijing, China

Patrick Girard
Department of Philosophy
The University of Auckland
Auckland, New Zealand

This paper presents a two-dimensional modal logic for reasoning about the changing patterns of knowledge and social relationships in networks organised on the basis of a symmetric 'friendship' relation, providing a precise language for exploring 'logic in the community' [11]. Agents are placed in the model, allowing us to express such indexical facts as 'I am your friend' and 'You, my friends, are in danger'.

The technical framework for this work is general dynamic dynamic logic (GDDL) [4], which provides a general method for extending modal logics with dynamic operators for reasoning about a wide range of model-transformations, starting with those definable in propositional dynamic logic (PDL) and extended to allow for the more subtle operators involved in, for example, private communication, as represented in dynamic epistemic logic (DEL) and related systems. We provide a hands-on introduction to GDDL, introducing elements of the formalism as we go, but leave the reader to consult [4] for technical details.

Instead, the purpose of this paper is to investigate a number of conceptual issues that arise when considering communication between agents in such networks, both from one agent to another, and broadcasts to socially-defined groups of agents, such as the group of my friends. All three components of the communication (the sender, the message, and the receivers) can be specified in a variety of ways that need to be distinguished. For example, Charlie may tell Bella 'you are in danger' or 'I am in danger'. He may broadcast to all 'my friends are in danger', which if Bella is a friend, will mean that that she is in danger, or send a message only to his friends that they are in danger. All such possibilities, together with their epistemic consequences, will be examined.

We extend the treatment of announcements to questions, in which agents are taken to be sincere and cooperative interlocutors, and consider network structure changing operations such as adding and deleting friends (with the permission of other agents) and, finally, explore the effect of all this on the concept of common knowledge, which is more varied and rich in the social network setting.

These issues are illustrated by a number of examples about office gossip, cold-war spy networks and Facebook.

## 1. A LANGUAGE OF SOCIAL KNOWING

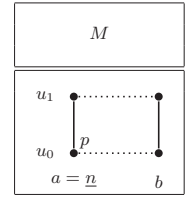

**Figure 1: A simple EFL model**

We start with a language $\mathcal{L}$ of *epistemic friendship logic* EFL based on atoms of two types: propositional variables $\rho \in \mathsf{Prop}$ representing indexical propositions such as 'I am in danger', and (a finite set of) agent nominals $n \in \mathsf{ANom}$ which stand for indexical propositions asserting identification: 'I am $n$'. The language is then inductively defined as:

$$\varphi ::= \rho \mid n \mid \neg\varphi \mid (\varphi \wedge \varphi) \mid K\varphi \mid F\varphi \mid A\varphi$$

We read $K$ as 'I know that' and $F$ as 'all my friends' and $A$ as 'every agent'. Models for this language are Kripke models of the form $M = \langle W, A, k, f, V \rangle$, where $W$ is a set (of epistemic states), $A$ is a set (of agents), and

1. $k$ is a family of equivalence relations $k_a$ for each agent $a \in A$, representing the ignorance of $a$ in distinguishing epistemic possibilities (as for standard S5 epistemic logic)

2. $f$ is a family of symmetric and irreflexive relations $f_w$ for each $w \in W$, representing the friendship relation in state $w$.

3. $g$ is a function mapping each agent nominal $n \in \mathsf{ANom}$ to the agent $g(n) \in A$ named by $n$. We abbreviate $g(n)$ to $\underline{n}$ when the model is clear from the context.

4. $V$ is a valuation function mapping propositional variables $\mathsf{Prop}$ to subsets of $W \times A$, with $(w, a) \in V(p)$ representing that the indexical proposition $p$ holds of agent $a$ in state $w$.

For example, Figure 1 illustrates a simple model for a language in which there is only one propositional variable $p$ and one agent name $n$. The set of states is $W = \{u_0, u_1\}$ and the set of agents is $A = \{a, b\}$, with $g(n) = a$, $n$ naming agent $a$. Both agents are ignorant about which state they are in, so $k_a = k_b$ is the universal relation. These are indicated by the two columns of the diagram. The left column displays the $k_a$ relation with a thick line; the right column displays the $k_b$ relation, similarly. The lines are non-directional because the relations are assumed to be symmetric. In more complex diagrams, we will assume that the relations depicted are the reflexive, transitive closures of what is shown explicitly. The rows of the diagram show the relations $f_{u_0}$ (first row) and $f_{u_1}$ (second row) with dotted lines. This represents the two agents being friends in both states of $W$. Again these are non-directional because we assume symmetry. But for these lines we do *not* take the reflexive, transitive closure, since we assume that $f_w$ is irreflexive and may or may not be transitive. Finally, that $p$ holds only of agent $a$ in state $u_0$, i.e., that $V(p) = \{(u_0, a)\}$ is shown by labelling the lower left node of the diagram with $p$.

Models are used to interpret $\mathcal{L}$ in a double-indexical way, as follows:

| | | |
|---|---|---|
| $M, w, a \models \rho$ | iff | $(w, a) \in V(\rho)$, for $\rho \in$ Prop |
| $M, w, a \models n$ | iff | $g(n) = a$, for $n \in$ ANom |
| $M, w, a \models \neg\varphi$ | iff | $M, w, a \not\models \varphi$ |
| $M, w, a \models (\varphi \wedge \psi)$ | iff | $M, w, a \models \varphi$ and $M, w, a \models \psi$ |
| $M, w, a \models K\varphi$ | iff | $M, v, a \models \varphi$ for every $v \in W$ such that $\langle w, v \rangle \in k_a(w)$ |
| $M, w, a \models F\varphi$ | iff | $M, w, b \models \varphi$ for every $b \in A$ such that $\langle a, b \rangle \in f_w(a)$ |
| $M, w, a \models A\varphi$ | iff | $M, w, b \models \varphi$ for every $b \in A$. |

As usual in modal logic, we can define the duals of the operators, which we write inside angle brackets: $\langle K \rangle = \neg K \neg$ 'it is epistemically possible for me that', $\langle F \rangle = \neg F \neg$ 'I have a friend who', and $\langle A \rangle = \neg A \neg$ 'there is someone who'. The English glosses are not so exact and require some manipulation to get proper translations, because of the way pronouns work in English. For example, if $d$ represents 'I am in danger' then $\langle F \rangle Kd$ means 'I have a friend who knows that he is in danger' rather than 'I have a friend who I know that I am in danger' which is not even grammatically correct.

We also use abbreviations for the hybrid-logic-like operators $@_n\varphi = A(n \rightarrow \varphi)$ (equivalently, $\langle A \rangle (n \wedge \varphi)$).[1] So, for example, if $\underline{n}$ is Charlie then the operator $@_n$ simply shifts the indexical subject to Charlie, so that $@_n d$ means 'Charlie is in danger'.

We say that $M$ is a *named agent* model, if every agent in $M$ has a name, i.e., for each $a \in A$, there is an $n \in$ ANom such that $g(n) = a$. The model depicted in Figure 1 is *not* a named agent model because agent $b$ has no name. In what follows we will assume that all agents are named, and so use the letters representing the agents in the diagram also as names in the language, abusing the distinction between $n$ and $\underline{n}$.

The advantage of working with named agent models is that we can define an operator $\downarrow n$ by

$$\downarrow n \; \varphi \quad := \quad \bigvee_{m \in \text{ANom}} (m \wedge \varphi[^n_m])$$

where $\varphi[^n_m]$ is the result of replacing agent nominal $n$ by $m$

in $\varphi$. This provides a way of referring to 'me' inside the scope of other operators, by shifting the referent of $n$ to the current agent. When $M$ is a named agent model,

$$M, w, a \models \downarrow n \; \varphi \qquad \text{iff} \quad M[^n_a], w, a \models \varphi.$$

where $M[^n_a]$ is the result of changing $M$ so that $n$ now names $a$.[2] This allows us to express such propositions as, $\downarrow n \, FK\langle F \rangle n$, which says 'all my friends know they are friends with me', at least on the assumption that every agent has a name. The assumption is not so restrictive, since in all applications we have so far considered, we can assume that a finite set of agents is specified in advance.[3]

*Relations and change.*

We will define a class of operators $\mathcal{D}$ and corresponding actions on models such that for each $\Delta \in \mathcal{D}$ and each $M$ model for $\mathcal{L}$, there is an $\mathcal{L}$ model $\Delta M$, and for each state $w$ of $M$, a state $\Delta w$ of $\Delta M$. We then extend $\mathcal{L}$ to a language $\mathcal{L}(\mathcal{D})$ of *dynamic epistemic friendship logic* (DEFL) by adding the elements of $\mathcal{D}$ as propositional operators and defining

$$M, w, a \models \Delta\varphi \qquad \text{iff} \quad \Delta M, \Delta w, a \models \varphi$$

To define $\mathcal{D}$, we use the language of propositional dynamic logic (PDL) with basic programs $K$, $F$ and $A$, given by

$$\begin{aligned} \mathcal{T} & \quad \pi ::= K \mid F \mid A \mid \varphi? \mid (\pi; \pi) \mid (\pi \cup \pi) \mid \pi^* \\ \mathcal{F} & \quad \varphi ::= \rho \mid n \mid \neg\varphi \mid (\varphi \vee \varphi) \mid \langle \pi \rangle \varphi \end{aligned}$$

for $\rho \in$ Prop and $n \in$ ANom. The denotation of program terms $\pi \in \mathcal{T}$ and formulas $\varphi \in \mathcal{F}$ in a model $M$ are defined in the manner shown in Table 1. Note in particular, the clauses for $K$, $F$ and $A$, in which these program terms refer to the accessibility relations of the corresponding operators of EFL, when interpreted two-dimensionally. Complex program terms are built up in the usual way: $(\pi_1; \pi_2)$ for the relational composition of $\pi_1$ and $\pi_2$, $(\pi_1 \cup \pi_2)$ for their union (or choice), $\varphi?$ for the 'test' consisting of a link from $(w, a)$ to itself iff $M, w, a \models \varphi$, and $\pi^*$ for the reflexive, transitive closure of $\pi$, which is understood as a form of iteration. Note also that we have abused notation so that formulas $\varphi$ of EFL, written with existential operators $\langle K \rangle$, $\langle F \rangle$ and $\langle A \rangle$, are also programs formulas (in $F$). This is justified by the obvious semantic equivalence:

$$M, w, a \models \varphi \text{ iff } (w, a) \in [\![\varphi]\!]^M$$

| | | |
|---|---|---|
| $[\![\rho]\!]^M$ | = | $V(\rho)$, for $\rho \in$ Prop |
| $[\![n]\!]^M$ | = | $W \times \{g(n)\}$, for $n \in$ ANom |
| $[\![(\varphi \wedge \psi)]\!]^M$ | = | $[\![\varphi]\!]^M \cap [\![\psi]\!]^M$ |
| $[\![\neg\varphi]\!]^M$ | = | $W \setminus [\![\varphi]\!]^M$ |
| $[\![\langle\pi\rangle\varphi]\!]^M$ | = | $\{w \in W \mid w[\![\pi]\!]^M v$ and $v \in [\![\varphi]\!]^M$ for some $v \in W \}$ |
| $[\![K]\!]^M$ | = | $\{\langle(w, a), (v, a)\rangle \mid k_a(w, v)\}$ |
| $[\![F]\!]^M$ | = | $\{\langle(w, a), (w, b)\rangle \mid f_w(a, b)\}$ |
| $[\![A]\!]^M$ | = | $\{\langle(w, a), (w, b)\rangle \mid a, b \in A, w \in W\}$ |
| $[\![\varphi?]\!]^M$ | = | $\{\langle w, w \rangle \mid w \in [\![\varphi]\!]^M\}$ |
| $[\![\pi_1; \pi_2]\!]^M$ | = | $\{\langle w, v \rangle \mid w[\![\pi_1]\!]^M s$ and $s[\![\pi_2]\!]^M v$ for some $s \in W\}$ |
| $[\![\pi_1 \cup \pi_2]\!]^M$ | = | $[\![\pi_1]\!]^M \cup [\![\pi_2]\!]^M$ |
| $[\![\pi^*]\!]^M$ | = | $\{\langle w, v \rangle | w = v$ or $w_i[\![\pi]\!]^M w_{i+1}$ for some $n \geq 0, w_0, \ldots, w_n \in W$, $w_0 = w$ and $w_n = v\}$ |

Table 1: Semantics of PDL terms and formulas

---

[1] Although reminiscent of hybrid logic, the 'agent nominals' $n$, binder $\downarrow n$ and now the operator $@_n$ are not exactly the same as their hybrid-logic namesakes, but are rather some sort of two-dimensional cousins. A true nominal, for example, is a proposition that is logically compelled to be satisfied by exactly one evaluation index, which in the case of our models, would have to be the pair $\langle w, a \rangle$.

[2] More precisely, $M[^n_a] = \langle W \times A, k, f, g[^n_a], V \rangle$ where for $m \in$ ANom, $g[^n_a](m) = a$ if $m = n$ and $g(m)$, otherwise.

[3] $\downarrow n$ can be introduced as a primitive, but without the restriction to named agent models, the resulting logic can be shown to be undecidable by encoding tiling problems (in the manner of [2]).

Now the class of dynamic operators will be defined using the theory of General Dynamic Dynamic Logic (GDDL) given in [4], which applies to any language of PDL. We refer the reader to that paper for full technical details, but we will introduce those parts of the theory that are required for present purposes.

The simplest GDDL operators are called PDL-*transformations*. These consist of assignment statements which transform models by redefining the basic programs. For example, the operator $[K := \pi]$ acts on model $M$ to produce a new model $[K := \pi]M$ such that

$$[\![K]\!]^{[K:=\pi]M} = [\![\pi]\!]^M$$

On states, there is no change: $[K := \pi]w = w$, so the resulting DEFL operator has the following semantics:

$$M, w, a \models [K{:=}\pi]\varphi \quad \text{iff} \quad [K{:=}\pi]M, w, a \models \varphi$$

We must be a little careful in the choice of $\pi$ so as to ensure that the resulting model $[K := \pi]M$ is still a model for EFL. For example, consider the program term $n?; K$. In $M$, this relates $(u_0, a)$ to $(u_1, b)$ in case $(u_0, a) \in [\![n]\!]^M$ and $(u_0, a)[\![K]\!]^M(u_1, b)$, which only holds when $g(n) = a$, $a = b$, and $k_a(u_0, u_1)$. Then $[K := n?; K]M$ is the structure $\langle W, A, k', f, V \rangle$ in which $k'_a = k_a$ and $k'_b = \emptyset$, for $b \neq a$. This is *not* a model for EFL. To make it into a model for EFL, we need to make each $k_a$ reflexive. This can be done with the program term $\top?$, since $[\![\top?]\!]^M$ is the identity relation. Thus taking $\pi$ to be $(n?; K) \cup \top?$ we get the model $[\![K]\!]^{[K:=(n?;K)\cup\top?]M}$ which is the structure $\langle W, A, k'', f, V \rangle$ in which $k''_a = k_a$ and $k''_b$ is the identity relation for all $b \neq a$. The application of $[K := (a?; K) \cup \top?]$ to a particular model is illustrated in Figure 2. Here, $M$ is a named agent model, so we allow ourselves the abuse of notation involved in writing $a$ for the name of $a$. In this model there are two friends, $a$ and $b$, who are both ignorant about whether they are in state $u_0$ or $u_1$. $p$ holds only of agent $a$ in state $u_0$, so in particular, $M, u_0, b \models (K \neg p \wedge \neg K \langle F \rangle p)$, which means that agent $b$ knows that she is not $p$ but does not know whether she has a friend who is $p$. After the action $[K := (n?; K) \cup \top?]$ we get the model shown on the right, in which $k_a$ is as before but now $k_b$ is the identity relation. In the transformed model, agent $b$ now knows that she has a friend who is $p$. Thus we get the dynamic fact:

$$M, u_0, b \models [K := (n?; K) \cup \top?]K\langle F \rangle p$$

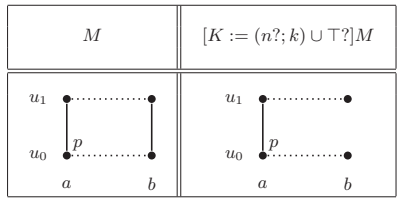

Figure 2: **A simple PDL-transformation.**

In effect, the PDL-transformation, $[K := (n?; K) \cup \top?]$ is the action of revealing everything to every agent other than $n$. We will consider more subtle forms of epistemic change in subsequent sections. Now it is time for a more extended example.

*The Spy Network.*
To take a Cold War example, suppose we are reasoning about the effect of a spy network being exposed.

> Bella ($b$) is friends with Charlie ($c$) and Erik ($e$), neither of whom are friends with each other. Unknown to the others is that Erik is a spy ($s$). The others are not spies, and Erik knows that because all spies know who else is a spy (we suppose). Bella knows that Charlie is not a spy, but Charlie does not know about her. After the network is exposed, all the spies and their friends will be interrogated by the police. But just before this happens a message is relayed to all agents revealing whether or not they are in danger, that is, whether they are a spy (which they would know in any case) or a friend of a spy.

A model $M$ of the initial situation is depicted in Figure 3, with $u_0$ representing the actual state. In EFL we can state pertinent facts such as $@_b(K \neg s \wedge \neg K \langle F \rangle s)$ 'Bella knows that she is not a spy but doesn't know if a friend of hers is a spy'. We will write $d$ 'I am in danger' as an abbreviation for $(s \vee \langle F \rangle s)$ 'either I'm a spy or I have a spy as a friend', and, for convenience, we have labelled those state-agent pairs at which $d$ holds. Thus we can read that $@_b(d \wedge \neg Kd)$ 'Bella is in danger but doesn't know it', whereas $@_bK@_c\neg d$ 'Bella knows that Charlie is not in danger'.

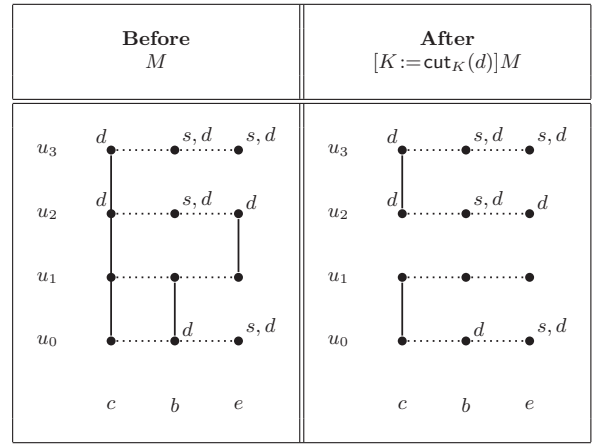

Figure 3: **Spy Network**

Now consider the PDL-term $\mathsf{cut}_K(\varphi)$ defined by

$$(\varphi?; K; \varphi?) \cup (\neg\varphi?; K; \neg\varphi?)$$

This relates $\langle w, a \rangle$ to $\langle v, b \rangle$ iff $a = b$, $k_a(w, v)$, and either $\varphi$ is true of $a$ in both states $w$ and $v$ or false of $a$ in both states. Thus the operator $[K := \mathsf{cut}_K(\varphi)]$ produces a new model $[K := \mathsf{cut}_K(\varphi)]M$ from $M$ by removing the $k_a$ links between states with conflicting values for $\varphi$ (about $a$). Effectively, this 'reveals' to each agent whether or not $\varphi$ holds (for them). This operator was first introduced in [14].
In our example, the situation after the revelation of $d$ 'you are in danger' is given by the model $[K := \mathsf{cut}_K(d)]M$, shown in the right part of Figure 3. Notice that the $k_c$ link between $u_1$ and $u_2$ are cut because $M, u_1, c \not\models d$ but $M, u_2, c \models d$; Charlie finds out that he is not in danger. Similarly, the $k_b$ link between $u_0$ and $u_1$ is cut because Bella finds out that she *is* in danger ($@_b Kd$). Finally, the $k_e$ link between $u_1$ and $u_2$ is cut because everyone now knows that Erik knows whether he is in danger (although only Bella knows which). Moreover, in the language of DEFL we can represent reasoning about these changes, such as the valid schema

$$[K := \mathsf{cut}_K(\varphi)]A(K\varphi \vee K\neg\varphi)$$

which states (for non-epistemic facts $\varphi$ such as $d = \langle F \rangle s$) that after $\varphi$ is revealed, everyone knows whether $\varphi$ or not.

GDDL *operators.*
More complicated operators can be constructed from finite relational structures whose elements are each associated with a PDL transformation, and whose combined effect on the a model is calculated by 'integrating' them according to a further such transformation. A GDDL operator $\Delta$ is something that looks like this:

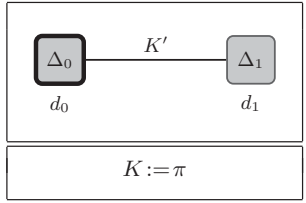

This represents an action $d_0$ (highlighted as the action that is actual performed) whose effect on the model is given by the PDL-transformation $\Delta_0$. There is also an action $d_1$ with associated PDL-transformation $\Delta_1$, and the relationship between $d_0$ and $d_1$ is marked as $K'$.[4] The effect of the operator on an EFL model $M$ with domain $W$ is computed by forming a product model $M'$ (in the manner of [1]) whose domain is $W \times \{d_0, d_1\}$, in which the elements $(w, d_i)$ represent the state resulting from action $d_0$ when the initial state is $w$. The model $M'$ consists of copies of two models $[\Delta_0]M$ with domain $W \times \{d_0\}$ and $[\Delta_1]M$ with domain $W \times \{d_1\}$, and a duplication of the model occurring in $\Delta$ itself, with, in this case, $(w, d_0)[\![K']\!]^{M'}(w, d_1)$ for each $w, v \in W$. Finally, the model $[\Delta]M$ is computed by applying the 'integrating' transformation $[K := \pi]$ to $M'$. This uses a PDL program term $\pi$ to compute the new value for $K$ from a combination of relations in the copied models $[\Delta_0]M$ and $[\Delta_1]M$ and the new relation $K'$ from $\Delta$ itself.[5]
This somewhat complex operation is best explained by looking at a simple example. Consider the case in which $\Delta_0$ is the PDL transformation $[K := (a?; K) \cup \top?]$ considered earlier, and $\Delta_1$ is the identity transformation, $I$. We will also take $\pi$ to be $(K \cup a?; K')^*$.

---
[4]In the general case, as explained in [4], there may be many actions and many new relation symbols; also, propositional variables.

[5]Again, the general case is more flexible, allowing any of the basic expressions $K, F$, agent nominal and propositional variable to be reinterpreted at the integrating stage.

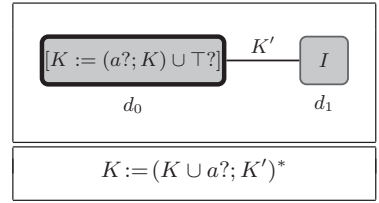

The action of this GDDL operator on the model $M$ considered earlier, is show in Figure 4. It represents a situation in which an action $d_0$ gives complete information to all agents other than $a$. The occurrence of $d_0$ is known to all agents other than $a$, who stays completely in the dark. Not only is $k_a$ unchanged in both $[\Delta_0]M$ (the top half of the diagram) and $[\Delta_1]M$ (the bottom half), but $a$ is also ignorant about which of these two submodels she is in, as represented by the vertical lines in connecting the two halves of the $a$ column: $(w, d_0)k_a(w, d_1)$ for all $w \in W$. Once again, we must check that the resulting model is an EFL model. In this case, it is. The $k_a$ and $k_b$ relations are transitive thanks to the application of the $*$ operator in the integrating transformation $[K := (K \cup a?; K')^*]$.
We'll say that a GDDL transformation $\Delta$ is a *general EFL dynamic operator* if it is in the language of PDL terms defined above, possibly augmented with internal relations such as $K'$ and also preserves the property of being a EFL-model: whenever $M$ is a EFL-model, so is $\Delta M$.

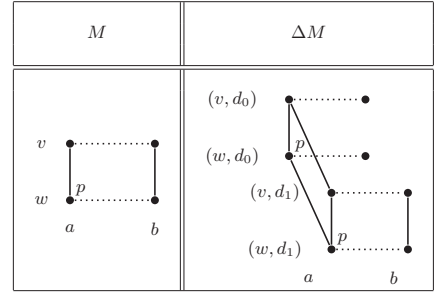

**Figure 4: A simple GDDL operator in action.**

## 2. SOCIAL ANNOUNCEMENTS

We now turn to direct communications, or 'announcements', within a social network. In the standard analysis of public announcement (PAL [7]), only the effect of announcement is modelled without reference to the agent who made the announcement and with the simplifying assumption that the message is received by all agents. In dynamic epistemic logic (following [1]), private announcements, in which a message is received by a limited set of agents are also considered. In the general case, within a social network, an announcement consists of an agent (the sender) transmitting some information (the message) to one or more other agents (the receivers) and each of these three components can be described in different ways, from different perspectives.[6] In this section, we will map out some of the subtleties.

---
[6]We are aware of the attempts by others in this respect. [8]

As a starting point, we ignore the sender and define a basic act of communication in which a message $\psi$ is sent (anonymously, we suppose) to a group of agents $\theta$ by

$$\mathsf{send}_\theta(\psi) = [K := (\theta?; \mathsf{cut}_K(\psi)) \cup (\neg\theta?; K)]$$

The action $\mathsf{send}_\theta(\psi)$ reveals the truth or falsity of $\psi$ (which may be different for different agents) to all agents satisfying $\theta$, and leaves the $k_a$ relation unchanged for agents $a$ not satisfying $\theta$.

To see how this works, consider $\mathsf{send}_{\langle F \rangle b}(d)$ in the case of our spy network. This is an anonymous announcement to the friends of Bella (but not to Bella herself) whether or not they are in danger. The effect of this action is shown in Figure 5. The formula $\theta$ describing the receivers of the message is $\langle F \rangle b$, which is satisfied by Charlie and Erik in the actual state $u_0$. Thus only the relations $k_c$ and $k_e$ are changed; $k_b$ remains the same. This is by no means our final analysis of communication. For one thing, actions of this sort are only 'semi-private', i.e., directed at particular individuals, but with others not involved in the communication still aware that it has occurred. Later, we will need to make the analysis more complex to cope with a great degree of privacy, in which only the sender and receivers are aware that the communication has occurred. For example, after the communication to Bella's friends, Bella knows something that she didn't know before: before she knew that Charlie was not in danger, now she knows that Charlie knows this:

$$M, u_0, b \models [\mathsf{send}_{\langle F \rangle b}(d)] K@_c K \neg d$$

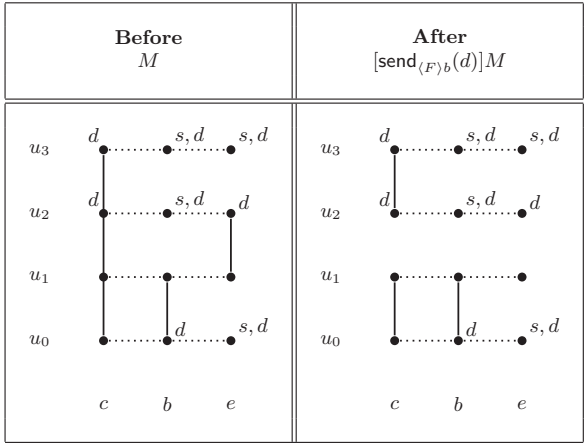

Figure 5: Restricting to Bella's friends

Yet before we get to the issue of privacy, we will bring the sender into our model, and explore some subtle distinctions about the nature of the message itself.

analysed specific types of communication network (i.e., communications that take place between one agent and another, or between an agent and a group of agents) when considering the issue of how distributed knowledge can be established by a group of agents through communication. Communication graphs were adopted by [6] to study communication between agents. Agent $i$ directly receiving information from agent $j$ is represented by an edge from agent $i$ to agent $j$ in such graph. Neither approach considers groups of agents described in terms of social relations.

*Announcements about the sender.*

The first case is that of a message sent by agent $n$ to agents described by $\theta$ with a message $\psi$, which is understood to be about the sender, for example 'I am in danger'. We define $[n \triangleleft \psi! : \theta]\varphi$, the statement that $\varphi$ holds such a communication, as

$$(@_n K\psi \to [\mathsf{send}_\theta(@_n \psi)]\varphi)$$

To make sense of this, we will look at a progression of simpler cases. First, with $\theta = \top$, the formula $[n \triangleleft \psi! : \top]\varphi$ means that $\varphi$ holds after agent $n$ publicly announces that $\psi$, noting that it simplifies to $(@_n K\psi \to [K := \mathsf{cut}_K(@_n \psi)]\varphi)$.

We make the rather strong assumption that the message is known by the sender.[7] Suppose, for example, that Erik, unable to keep his secret any longer, told everyone that he is a spy. After this, everyone would know that he is a spy (and Bella, his friend, would know that she is in danger). This follows from the validity of $[e \triangleleft s! : \top] AK@_e s$.[8] Note that $[b \triangleleft s! : \top] AK@_b s$ is also true (since it is valid!). This says that everyone would know that Bella is a spy after she announced it. But the reason is quite different: Bella could not announce that she is a spy, because she knows that she isn't.[9]

The second case is an announcement to a particular agent. In this case, $\theta$ is an agent nominal $m$ and the formula $[n \triangleleft \psi! : m]\varphi$ means that $\varphi$ holds after agent $n$ announces to $m$ that $\psi$. For example, Erik may be more cautious in his admission, telling only Bella, after which she, but not Charlie would know: $[e \triangleleft s! : b] @_b K@_e s$ and $(\neg(b \lor K@_e s) \to [e \triangleleft s! : b] \neg K@_e s)$ are both valid, and the latter says that an agent who is neither Bella nor (already) knows that Erik is a spy, still doesn't know this after he announces it to Bella. In the most general case, $\theta$ is a description of a group of agents. For example, $[b \triangleleft \neg s! : \langle F \rangle b]\varphi$ states that $\varphi$ would hold after Bella tells her friends that she is not a spy. Again we have a useful validity: $[b \triangleleft \neg s! : \langle F \rangle b] @_b F K@_{b} \neg s$, which says that if Bella were to tell her friends that she is not a spy then they would all know that she isn't a spy.

*Announcements about the receivers.*

Announcements that are indexical about the receiver such as 'you are in danger' (announced to Bella by Erik) or 'you are my friends' (announced by Bella to her friends) can be expressed with a slight change that captures the different preconditions for announcements. We define $[n : \psi! \triangleright \theta]\varphi$, the statement that $\varphi$ holds after agent $n$ announces message $\psi$ (about $\theta$) to agents satisfying $\theta$ as

$$(@_n A(\theta \to \psi) \to [\mathsf{send}_\theta(\psi)]\varphi)$$

Again, we first consider the simple case of public announcement, represented by $[n : \psi! \triangleright \top]\varphi$, which can be seen to be equivalent to $(@_n KA\psi \to [K := \mathsf{cut}_K(\psi)]\varphi)$. Consider, for example, my announcing to everyone 'you are in danger'.

---

[7]The standard assumption of PAL that announcements are true is thus equivalent to supposing that they are made by God, or some other omniscient entity. [5] studied different types of agent (truth-teller, liar and bluffer), how they make announcements, and are subsequently interpreted in communication.

[8]In fact, the information that Erik is a spy becomes common knowledge, as we will see in Section 6.

[9]It would be enough for Bella merely not to know that she is a spy for the announcement to be impossible.

The precondition that I know everyone is in danger is captured by the antecedent $KAd$, and after the announcement everyone knows that she is in danger, as is represented by the validity of $\downarrow n\ [n\!:\!d!\triangleright\top]AKd$.

The case of agent-to-agent announcement displays a nice symmetry between the two kinds of indexical message. Agent $n$ announcing 'you are in danger' to agent $m$ is equivalent to announcing (again to $m$) that $m$ is in danger. More generally, the following equivalences are valid

$$[n\!:\!\psi!\triangleright m]\varphi \ \leftrightarrow\ [n\triangleleft @_m\psi!\!:\!m]\varphi$$
$$[n\triangleleft\psi!\!:\!m]\varphi \ \leftrightarrow\ [n\!:\!@_n\psi!\triangleright m]\varphi$$

This symmetry between announcements is more delicate when announcing to groups. Announcing 'you are in danger' to each of my friends is only the same as announcing to them 'all my friends are in danger' on the assumption that each friend knows only that she is my friend, and knows nothing about the others. Without this assumption,

$$[n\!:\!\psi!\triangleright\langle F\rangle n]\varphi \ \leftrightarrow\ [n\triangleleft @_n F\psi!\!:\!\langle F\rangle n]\varphi$$

is not always valid.[10]

For announcement to friends, an interesting new phenomenon arises. Consider the case of my announcing 'you are my friend' to my friends. That $\varphi$ holds after such an announcement is represented by $[n\!:\!\langle F\rangle n!\triangleright\langle F\rangle n]$. The message is the same as the description of the set of receivers, so when this is expanded, we find that the precondition for the announcement is $\downarrow n\ KA(\langle F\rangle n\to\langle F\rangle n)$, which is valid, so the announcement can always be made, by anyone. But nonetheless, it can be informative, as can be seen from the validity of $\downarrow n\ [n\!:\!\langle F\rangle n!\triangleright\langle F\rangle n]FK\langle F\rangle n$, which says that after my making this announcement, my friends all know that they are my friends, something they may not have known before.

Finally, we note that any sender-indexical announcement to a group $\theta$ is equivalent to a receiver-indexical announcement to the same group $\theta$ in the case that there is at least one receiver ($A\neg\theta$ is false). The trick is that the statement $\psi$ about $n$ (the sender) is then equivalent to the statement $@_n\psi$ about any (every) receiver. More formally, the following is valid:[11]

$$(\neg A\neg\theta \to [n\triangleleft\psi!\!:\!\theta]\varphi \ \leftrightarrow\ [n\!:\!@_n\psi!\triangleright\theta]\varphi)$$

*Private announcements.*
Communications of the form $[n\triangleleft\psi!\!:\!\theta]$ and $[n\!:\!\psi!\triangleright\theta]$ are only semi-private. Their effect on the model ensures that

---

[10] For a simple counterexample, consider $\psi$ to be $d$ and the model $M$ (shown left).

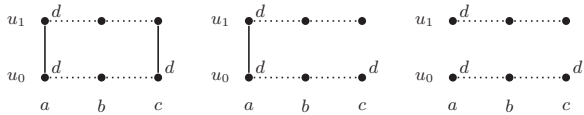

The precondition of $[b\!:\!d!\triangleright\langle F\rangle b]$ is $@_b KA(\langle F\rangle b\to d)$, which is equivalent to the precondition $@_b KFd$ of $[b\triangleleft @_b Fd!\!:\!\langle F\rangle b]$ which is satisfied in $M$, and the resulting two models are shown middle and right. Yet these are easily distinguished, by taking $\varphi$ to be $@_a K@_c d$.

[11] The key observation here is that the precondition for the sender-indexical announcement is $@_n K\psi$, which is equivalent to the precondition $@_n KU_A(\theta\to @_n\psi)$ when $U_A\neg\theta$ is false.

every agent will know that the announcement has occurred, if the sender satisfies the precondition, so, for example,

$$\downarrow n\ [n\triangleleft d!\!:\!m]AK(@_n Kd\to @_m K@_n d)$$

is valid: after I announce to $m$ that I am in danger, everyone will know that if I know I am in danger then $m$ also knows it. This is (typically) an unjustified violation of the privacy of the communication between me and $m$.

To make the action $\mathsf{send}_\theta(\psi)$ private, we embed it in a GDDL operator similar to the one given in our earlier example. Thus, for the sender-indexical[12] version, that $\varphi$ would hold after the private announcement of $\psi$ by $n$ to agents $\theta$ is be represented as

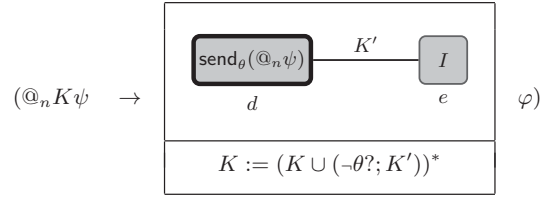

Call this formula $[\![n\triangleleft\psi!\!:\!\theta]\!]\varphi$. Inside the GDDL operator, the internal relation $K'$ represents ignorance about whether the communication $\mathsf{send}_\theta^n(\psi)$ has occurred or not, the latter possibility represented by the identity transformation, $I$. The integrating transformation $[K:=(K\cup(\neg\theta?;K'))^*]$ restricts ignorance of the $K'$ kind to agents other than $\theta$ and factors this in to the new epistemic relation. The $*$ is needed to ensure that the result is an equivalence relation. We will see an example of this operator in action at the end of the next section.

## 3. KNOWING YOUR FRIENDS

So far, the friendship relation in our models has been relatively tame, remaining fixed across epistemic states. We have used it to determine which group of agents receive a message, and even to specify the content of a message, but we have not yet considered ignorance about who is friends with whom. This is where it gets really interesting. We will explore some of the possibilities with an everyday example of infidelity and gossip.

> Peggy ($p$) knows that Roger ($r$) is cheating ($c$) on his wife, Mona ($m$). What's more, Roger knows that Peggy knows, because they met accidentally while he was with his mistress. Mona does not know about the affair, and both Peggy and Roger know this. The situation (for Roger) deteriorates when he discovers that Peggy is a terrible gossip. She is bound to have told all her friends about his affair. What Roger does not know is whether Mona is a friend of Peggy (she is).

We can represent the epistemic state of this network before Peggy's announcement with the model depicted in Figure 6, assuming that married couples are also friends. (The grey construction lines are only included to make the diagram easier to read; they have no epistemic or social significance.)

---

[12] The receiver-indexical version is obtained by changing the message and the precondition as in the simple semi-private case.

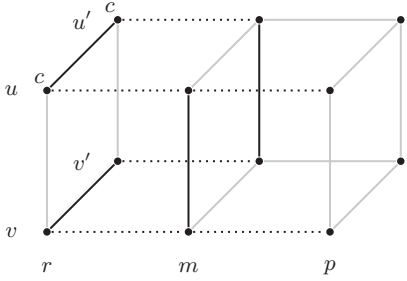

Figure 6: Roger's Quandry

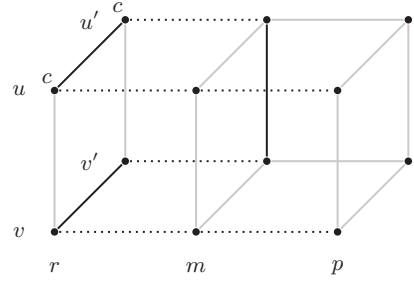

Figure 7: After Peggy's gossip

Note that the friendship relations are now different in different states. At $u$ (the actual state) for Roger $r$, the statements listed in Table 2 are all true. As a result, we can compute that at $w$ in the original model for Roger $r$, the formula

$$\downarrow n\ [p \triangleleft @_n c!\colon \langle F \rangle p] @_m K @_n c$$

is true, i.e., "I don't know that Mona will know about my cheating after Peggy tells her friends about it." That some proposition $\varphi$ holds after the announcement 'Roger is cheating!' that Peggy makes to her friends is given by $[p \triangleleft @_r c!\colon \langle F \rangle p] \varphi$, which expands and simplifies to

$$(@_p K @_r c \to [K := (\langle F \rangle p?;\mathsf{cut}_K(@_r c)) \cup (\neg \langle F \rangle p?; K)] \varphi)$$

| $c$ | I'm cheating |
|---|---|
| $\downarrow n\ K(@_p K @_n c \wedge @_m \neg K @_n c)$ | I know that Peggy (but not Mona) knows I am cheating. |
| $\downarrow n\ @_p K @_n K @_p K @_n c$ | Peggy knows I know she knows I am cheating |
| $\neg K @_m \langle F \rangle p \wedge \neg K @_m \neg \langle F \rangle p$ | I don't know whether Peggy and Mona are friends. |
| $\downarrow n\ @_p K @_n \neg K @_m \langle F \rangle p$ | Peggy knows I don't know whether she and Mona are friends. |

Table 2: Facts about Roger

When evaluated at $u$, the presupposition that Peggy knows that Roger is cheating is satisfied, and so the formula $\varphi$ is evaluated in the transformed model shown in Figure 7. (Note the missing vertical line in the middle.)
This is all very well, but Roger needs a little more privacy.

> Before returning home to face Mona, Roger is uneasy. He would really like to know whether or not she knows about his affair. He already knows that she knows if and only if she is friends with Peggy. So if Peggy told him that they are friends, he would be prepared for Mona's fury. But for his planned excuses to be convincing, Mona must not know that he knows she knows (about the affair). It is therefore very important that Peggy tells him in private.

Now let us suppose that the ever-loquacious Peggy announces to Robert privately that Mona is her friend, represented as $[\![p \triangleleft \langle F \rangle m!\colon r]\!]$. Now, whether the crucial proposition $\varphi$

$$(@_r K @_m K @_r c \wedge \neg @_m K @_r K @_m K @_r c)$$

(that Roger knows Mona knows he has been cheating but Mona doesn't know that he knows) holds must be determined by evaluating it in the model obtained by transforming the one in Figure 7 using the following GDDL operator, call it $\Delta$:

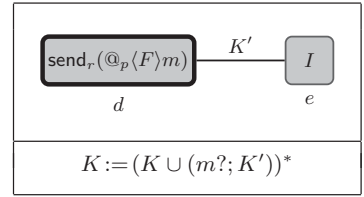

The result is shown in Figure 8.
The upper half of the diagram represent the result of action $d$, Peggy telling Roger that she is friends with Mona ($\mathsf{send}_r(@_p \langle F \rangle m)$), whereas the lower half represent the result of action $e$, nothing ($I$); it is just a copy of the model in Figure 7. Mona is the only one of the three who doesn't know which action has taken place, and her ignorance is represented by the lines connected corresponding states in the upper and lower halves (in the $m$ column). We see that $K @_m K @_r c$ holds of $r$ in state $(u, d)$, so Roger can meet Mona prepared.[13]
We may wonder about the accuracy of the model in representing Roger and Mona as friends after Peggy's announcement. Changes to the social network will be considered in Section 5.

## 4. ASKING QUESTIONS

As well as making announcements, agents in a social network can ask questions. Our approach to modelling questions will assume that agents are cooperative to the extent that they answer those questions to which they know the answer.[14] A more elaborate model would consider the preferences of agents, but that is beyond the scope of the current paper. With this assumption, the effect of asking whether $\psi$ of an agent $a$ who knows that $\psi$ is the same as an announcement by $a$ that $\psi$. Likewise, the effect of asking whether $\psi$ of an agent $a$ who knows that $\neg\psi$ is the same as an announcement by $a$ that $\neg\psi$. In the case that $a$ does not know whether $\psi$, we assume that this also is communicated (possibly by the mere absence of an expected reply). With this in mind, we define $[n \triangleleft \psi?\colon m]\varphi$, the proposition that $\varphi$ holds after agent $n$ asks agent $m$ whether $\psi$ as

$$([m \triangleleft \psi!\colon n]\varphi \wedge [m \triangleleft \neg\psi!\colon n]\varphi \wedge [m \triangleleft \neg(K\psi \vee K\neg\psi)!\colon n]\varphi$$

---
[13]Even the additional level of privacy offered here is still not perfect, as it involves some change in Mona's knowledge. She goes from knowing that Roger doesn't know that she is friends with Peggy to not knowing this. However, one may just think that privacy is a matter of degree.

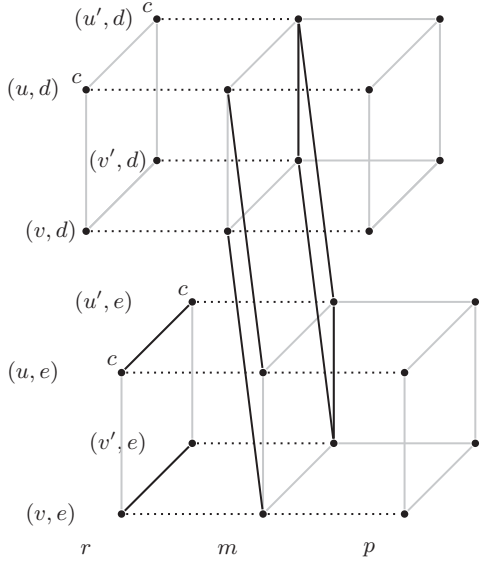

**Figure 8: Peggy to Roger, privately.**

In other words, $\varphi$ holds after $n$ asks $m$ whether $\psi$ just in case $\varphi$ holds after in all three cases: (1) $m$ answers 'yes', so announcing $\psi$ to $n$ (2) $m$ answers 'no', so announcing $\neg\psi$ to $n$ and (3) $m$ answers 'I don't know', so announcing $\neg(K\psi \vee K\neg\psi)$ to $n$. This ensures that the following are valid:

$$(@_m K @_n p \to [n \triangleleft p?\colon m]@_n Kp)$$
$$(@_m K @_n \neg p \to [n \triangleleft p?\colon m]@_n K\neg p)$$
$$(@_m \neg (K @_n p \vee K @_n \neg p)$$
$$\to [n \triangleleft p?\colon m]@_n K @_m \neg (K @_n p \vee K @_n \neg p))$$

So, for example, after Charlie $c$ asks Erik $e$ whether he (Charlie) is in danger, $d$, he will either know that he is in danger $Kd$ or know that he is not in danger $K\neg d$, or know that Erik doesn't know whether or not he (Charlie) is in danger, $\downarrow n\ K@_e \neg (K@_n d \vee K@_n \neg d)$.

Sender-indexical questions can be distinguished from receiver-indexical questions in a way that parallels the distinction for announcements. The question 'Are you in danger?' from $n$ to $m$, answered positively amounts to an announcement by $m$ to $n$ of 'I am in danger', and similarly with the 'you' and 'I' reversed.

As with announcements, this model of questions assumes that the answers are only semi-private. For example, after Charlie asks Erik whether he is in danger, a third-party will know that Charlie either knows whether he is in danger or knows that Erik doesn't know the answer. To make questioning more private, we need private announcements too. Here we will give one simple example.

> Roger approaches Peggy in private and asks her directly whether or not she and Mona are friends. Being sincere and cooperative, Peggy answers that they are. Mona, of course, knows nothing of their conversation.

This private question $[\![r\colon \langle F\rangle m?\colon p]\!]$ is defined by direct analogy with the semi-private question $[r \triangleleft \langle F\rangle m?\colon p]$ so that $\varphi$ holds after the question is asked just in case

$$[\![p \triangleleft \langle F\rangle m!\colon r]\!]\varphi \wedge [\![p \triangleleft \neg \langle F\rangle m!\colon r]\!]\varphi \wedge [\![p \triangleleft \neg(K\langle F\rangle m \vee K\neg \langle F\rangle m)!\colon r]\!]\varphi$$

In this case, only the precondition of $[\![p \triangleleft \langle F\rangle m!\colon r]\!]$ is satisfied, and so the results are just as depicted in Figure 8. Questions to groups present some further challenges. How would sincere and cooperative friends answer the question 'Am I in danger?'? For our present strategy to work they would have to do so by making an announcement. The problem is that if I have more than one friend who knows the answer, more than one announcement will follow. But in which order? Clearly, we must consider all possible orders, which in the general case involves quantification over an arbitrary number of friends. In finite named agent models this is possible, but a bit ugly, so we will pass over the details here.

## 5. CHANGING THE NETWORK

What makes networking intriguing is the dynamics of network changes. You can be friends with someone one day on Facebook, but you may drop him as a friend the following day or add someone else. Those acts, though simple, have a direct impact on information flow in communities. Consider the following:

> Roger, scared of the possibility that Mona will find out about his affair from Peggy, does all that he can to distance them. His smear campaign is designed to break their friendship and so protect his information.

To define the operation of deleting a friendship link, we first define the result of cutting the friendship link between agents $n$ and $m$ in one direction

$$\mathsf{cut}_F(n, m) = (\neg n?; F) \cup (F; \neg m?)$$

Then, to deleting the link between $n$ and $m$ we need to cut in both directions:[15]

$$[-F_{n,m}] = [F := \mathsf{cut}_F(n, m)][F := \mathsf{cut}_F(m, n)]$$

It is then fairly easy to show that $[\![F]\!]^{[-F_{nm}]M} = [\![F]\!]^M \setminus \{\langle n, m\rangle, \langle m, n\rangle\}$, as required.[16]

---

[14] For dealing with questions in terms of issue management in standard dynamic epistemic logic, we refer to [13]. Here we take a short-cut that reduces the action of asking a question to that of announcing the answer.

[15] It is also interesting to consider asymmetric relationships such as "following" on Twitter or "subscribing" on Facebook, as studied in [9].

[16] This follows from the fact that $a[\![F]\!]^{[F:=\mathsf{cut}_F(n,m)]M}b$ iff $a[\![F]\!]^M b$ and $\langle a, b\rangle \neq \langle n, m\rangle$.

Now how is this going to help Roger? Well, after the application of $[-F_{mp}]$ to the model of Figure **??**, Peggy's announcement to her friends that Roger is cheating has no effect; in fact, she has no friends to receive the message. So the model is unchanged. In other words, in this original model, it is true for Roger that

$$[-F_{mp}]\downarrow n\ [p \triangleleft @_n c! \colon \langle F\rangle p]@_m \neg K@_n c$$

'after Peggy loses Mona as a friend, even after she tells her friends that I am cheating, Mona won't know.'

Next we consider adding a friend. In the basic case, we can define the operation $[+F_{n,m}]$ by analogy with deletion, but more simply, as

$$[F =: F \cup (n?; A; m?)]$$

But a more interesting model of adding friends follows the protocol of Facebook and other online social networks, whereby one must first request friendship. To capture this aspect of network change, we need to represent whether or not an agent *wants* to be friends with another agent. In a fuller account, this could be done with a preference order, showing that the agent prefers states in which they are friends to those in which they are not. But for now, suppose that there is some additional indexical relation $d_w$ in our models, with $d_w(a, b)$ interpreted to mean that in state $w$, agent $a$ wants to become friends with agent $b$. Let $D$ be the corresponding modal operator.

The question 'do you want to be my friend?' from $n$ to $m$ is thus represented by $[n \triangleleft \langle D\rangle n? \colon m]$, but as a *request* we interpret this as involving an action: if the answer is 'yes' then we become friends; otherwise, there is no change to the social network, thought there are consequent epistemic changes, such as my learning that you don't want to be my friend. That $\varphi$ holds after this 'friend request' is therefore represented by

$$[\mathsf{add}(m)]\varphi = \ \downarrow n\ [n \triangleleft \langle D\rangle n? \colon m]((K@_m\langle D\rangle n \wedge [+F_{n,m}]\varphi) \\ \vee (\neg K@_m\langle D\rangle n \wedge \varphi))$$

A private version of this operation can be obtained by replacing the announcement and network change by a GDDL-based version.

The following validity shows some of the epistemic consequence of friend requests:

$\downarrow n\ ((\neg\langle F\rangle m \wedge \neg K@_m\langle D\rangle n) \to$
$[\mathsf{add}(m)]((K@_m K\langle D\rangle n \wedge \langle F\rangle m) \vee (K@_m \neg K\langle D\rangle n \wedge \neg\langle F\rangle m))$

If I'm not friends with $m$ and don't know that she wants to be my friend, then were I to ask her, I would either know that she knows she wants to be friends and we would be friends, or know that she doesn't know she wants to be friends and we wouldn't be friends.

## 6. COMMON KNOWLEDGE

In the context of social networks or communities, common knowledge is clearly an important notion. One can easily imagine the situations in which we want to reason about whether or not something is commonly known in some community or among my friends. There are at least two subtleties involved in making this precise. The first has to do with identifying the group of agents who are said to have common knowledge. This may be by means of a specific list ('Charlie, Bella, and Erik'), or a description ('Charlie's friends') or even an indexical description ('friends of mine'). Secondly, the information that is shared may be rigid ('it is common knowledge that Charlie is not a spy') or indexical (e.g. 'it is common knowledge among Charlie's friends that I am in danger' or 'it is common knowledge among my friends that they are in danger.')

To capture all these cases, first define $\overline{K}_a$ to be $(A; a?; K)$. Then $[\overline{K}_a]\varphi$ means that agent $a$ knows that $\varphi$, as justified by the following equivalence:

$M, w, b \models [\overline{K}_a]\varphi$   iff   $M, v, a \models \varphi$ for all $v \in W$ such that $k_a(w, v)$.

Here $\varphi$ could be an indexical proposition, so, for example, 'Charlie knows that he is not a spy' would be represented by $[\overline{K}_c]\neg s$, whereas 'Bella knows that Charlie is not a spy' would have to be represented as $[\overline{K}_b]@_c\neg s$. Now, for common knowledge, define

$$\mathsf{c}_\theta = (A; \theta?; K)^*; A; \theta?$$

and interpret $[\mathsf{c}_\theta]\varphi$ to mean, roughly, that there is common knowledge among $\theta$-agents that $\varphi$. So this enables us to talk, in our formal language, about the common knowledge of some group. This definition seems more general than the standard notion of common knowledge (see e.g. [3]). It is justified by the following applications, each of which can be suitably generalised.

1. Common knowledge among an enumerated set of agents about a non-indexical proposition. For example, that there is common knowledge between Bella ($b$) and Charlie ($c$) that Charlie is not a spy ($s$) can be represented by $[\mathsf{c}_{(b\vee c)}]@_c\neg s$.[17] To justify this claim, first note that the standard way of defining common knowledge for a group of agents $G$ is to introduce a new operator $C_G$ such that

    $M, w, a \models C_G\varphi$   iff   $M, v, a \models \varphi$ for all $\langle u, v\rangle \in (\bigcup_{a'\in G} k_{a'})^*$

    We can then prove that, for example, $[\mathsf{c}_{(b\vee c)}]@_c\neg s$ is equivalent to $C_{\{b,c\}}@_c\neg s$.[18]

2. Common knowledge among a non-indexically described group of agents about a non-indexical proposition. For example, that it is common knowledge among Peggy's ($p$) friends that Roger ($r$) is cheating ($c$) can be represented as $[\mathsf{c}_{\langle F\rangle p}]@_r c$. This implies that every friend of Peggy knows that Roger is cheating ($@_p FK@_r c$), but also that each of them knows that all of Peggy's friends know this ($@_p FK@_p FK@_r c$), and that each of them knows they all know *that* ($@_p FK@_p FK@_p FK@_r c$), and so on. As such, it is not equivalent to any statement of the form $C_G\varphi$. In particular, if, say, Peggy's only friends are Mona ($m$) and Nancy ($n$), it may

---

[17]Another concrete and interesting area of application is our ordinary email exchange, see an interesting analysis in [12].
[18]The argument is simple. First note that $(A; (b\vee c)?; K)^*$ is equivalent to $(\overline{K}_b \cup \overline{K}_c)^*$. Also, since $@_c\neg s$ is non-indexical, $[A; (b \vee c)?]@_c\neg s$ is equivalent to $@_c\neg s$. Thus $[\mathsf{c}_{(b\vee c)}]@_c\neg s$ is equivalent to $[(\overline{K}_b \cup \overline{K}_c)^*]@_c\neg s$, which is obviously equivalent to $C_{\{b,c\}}@_c\neg s$.

*not* have the same truth value as $C_{\{m,n\}}@_r c$, which is compatible with Mona's and Nancy's ignorance about what Peggy's friends (in general) know.

3. Common knowledge among a non-indexically described group of agents about a proposition that is indexical with respect to each member of the group. This is the subtlest case. For example, after the spy network has been exposed, that it is common knowledge among Erik's ($e$) friends that they are in danger ($d$) is represented by $[\mathsf{c}_{\langle F \rangle e}]d$. This implies that every friend of Erik (the spy) knows that s/he is in danger ($@_e FKd$), that each of them knows they all know this ($@_e FK@_e FKd$), and so on. Again, this is compatible with their ignorance about the friendship relation, so long as in all epistemically indistinguishable states, the friends of Erik (whoever they may be) are still in danger. The reason to have the final part $A; \theta$? in the above definition of $\mathsf{c}_\theta$ is this: when $\varphi$ is indexical, we need to ensure that it is about the members of $\theta$. When $\varphi$ is not indexical, this part is redundant.

4. Common knowledge among an indexically described group of agents about a non-indexical proposition. For example, that it is common knowledge among my friends that Roger is cheating is represented by $\downarrow n\ [\mathsf{c}_{\langle F \rangle n}]@_r c$. This is a straightforward generalisation of the previous case to an indexically specified description, with the $\langle F \rangle n$ using the nominal $n$, which is bound to the speaker by $\downarrow n$.

5. Common knowledge among an indexically described group of agents about a proposition that is indexical with respect to the speaker. For example, that there is common knowledge among my friends that I am not a spy is represented by $\downarrow n\ [\mathsf{c}_{\langle F \rangle n}]@_n \neg s$. This is really no more complicated than the last case. Again, the indexical work is all done by $\downarrow n$ in creating a temporary name 'n' for the speaker. Within that context, both the description of group ($\langle F \rangle n$) and the content of the common knowledge $@_n \neg s$ are both non-indexical.

6. Common knowledge among an indexically described group of agents about a proposition that is indexical with respect to each member of the group. For example, that it is common knowledge among my friends that they are in danger represented by $\downarrow n\ [\mathsf{c}_{\langle F \rangle n}]d$. This is an obvious generalisation of the previous cases.

Other useful specifications of groups of agents as the subjects of common knowledge include 'common knowledge of $\varphi$ in my community' ($\downarrow n\ [\mathsf{c}_{\langle f^* \rangle n}]\varphi$), 'common knowledge of $\varphi$ among those who know they are in danger' ($[\mathsf{c}_{Kd}]\varphi$), 'common knowledge of $\varphi$ among those who know they are my friends' ($\downarrow n\ [\mathsf{c}_{K \langle F \rangle n}]\varphi$).

## 7. CONCLUDING REMARKS

What has emerged from this study is an appreciation of the diversity of subtle logic distinctions when combining epistemic and social relations, especially when allowing indexical propositions, as are very common in the social setting. Although Facebook was an inspiration for this work, we have only scratched the surface. Facebook offers many interesting features that would be good to model, such as the wall, commenting, and liking. There are many directions in which the rather tight assumptions of epistemic friendship logic can be relaxed, such as by dropping symmetry for friendship, allowing degrees or hierarchies of friends (as in [10]), diluting knowledge to belief and adding preference.

## 8. REFERENCES

[1] A. Baltag, L. S. Moss, and S. Solecki. The logic of public announcements, common knowledge and private suspicious. Technical Report SEN-R9922, CWI, Amsterdam, 1999.

[2] P. Blackburn and J. Seligman. What are hybrid languages? In M. Kracht, M. de Rijke, H. Wansing, and M. Zakharyaschev, editors, *Advances in Modal Logic*, volume 1 of , pages 41–62. CSLI Publications, Stanford University, 1998.

[3] R. Fagin, J. Halpern, Y. Moses, and M. Vardi. *Reasoning about Knowledge*. The MIT Press, 1995.

[4] P. Girard, J. Seligman, and F. Liu. General dynamic dynamic logic. In T. Bolander, T. Braüner, S. Ghilardi, and L. S. Moss, editors, *Advances in Modal Logics Volume 9*, pages 239–260, 2012.

[5] F. Liu and Y. Wang. Reasoning about agent types and the hardest logic puzzle ever. *Minds and Machines*, 2013. To appear.

[6] E. Pacuit and R. Parikh. Reasoning about communication graphs. In D. G. Johan van Benthem, Benedikt Loewe, editor, *Interactive Logic*, pages 13–60. Amsterdam University Press, 2007.

[7] J. Plaza. Logics of public announcements. In *Proceedings of the 4th International Symposium on Methodologies for Intelligent Systems*, 1989.

[8] F. Roelofsen. Exploring logical perspectives on distributed information and its dynamics. Master's thesis, ILLC, The University of Amsterdam, 2005.

[9] J. Ruan and M. Thielscher. A logic for knowledge flow in social networks. In *Australasian Conference on Artificial Intelligence*, pages 511–520, 2011.

[10] J. Seligman, P. Girard, and F. Liu. Logical dynamics of belief change in the community. 2013. Under submission.

[11] J. Seligman, F. Liu, and P. Girard. Logic in the community. In M. Banerjee and A. Seth, editors, *ICLA*, volume 6521 of *Lecture Notes in Computer Science*, pages 178–188, 2011.

[12] F. Sietsma and K. Apt. Common knowledge in email exchanges. In J. Eijck and R. Verbrugge, editors, *Proceedings of the Workshop on Reasoning About Other Minds: Logical and Cognitive Perspectives*, pages 5–19, 2011.

[13] J. van Benthem and Ştefan Minică. Toward a dynamic logic of questions. In E. P. X. He, J. Horty, editor, *Logic Rationality and Interaction*, pages 27–41, Chongqing, China, August 2009. Springer.

[14] J. van Benthem and F. Liu. Dynamic logic of preference upgrade. *Journal of Applied Non-Classical Logic*, 17:157–182, 2007.